# Magnetotransport properties of the layered CaAl$_2$Si$_2$ semimetal hosting multiple nontrivial topological states

Hao Su[1,2,3,†], Xianbiao Shi[4,5,†], Wei Xia[1,2,3,†], Hongyuan Wang[1,2,3], Xuesong Hanli[1], Zhenhai Yu[1], Xia Wang[1,6], Zhiqiang Zou[1,6], Na Yu[1,6], Weiwei Zhao[4,5*], Gang Xu[7*], Yanfeng Guo[1*]

[1] School of Physical Science and Technology, ShanghaiTech University, Shanghai 201210, China
[2] Shanghai Institute of Optics and Fine Mechanics, Chinese Academy of Sciences, Shanghai 201800, China
[3] University of Chinese Academy of Sciences, Beijing 100049, China
[4] State Key Laboratory of Advanced Welding & Joining and Flexible Printed Electronics Technology Center, Harbin Institute of Technology, Shenzhen 518055, China
[5] Key Laboratory of Micro-systems and Micro-structures Manufacturing of Ministry of Education, Harbin Institute of Technology, Harbin 150001, China
[6] Analytical Instrumentation Center, School of Physical Science and Technology, ShanghaiTech University, Shanghai 201210, China
[7] Wuhan National High Magnetic Field Center and School of Physics, Huazhong University of Science and Technology, Wuhan 430074, China

Combination of different nontrivial topological states in a single material is capable of realizing multiple functionalities and exotic physics, but such materials are still very sparse. We report herein the results of magnetotransport measurements and *ab initio* calculations on single crystalline CaAl$_2$Si$_2$ semimetal. The transport properties could be well understood in connection with the two-band model, agreeing well with the theoretical calculations indicating four main sheets of Fermi surface consisting of three hole pockets centered at the Γ point and one electron pocket centered at the M point in the Brillouin zone. The single fundamental frequency imposed in the quantum oscillations of magnetoresistance corresponds to the electron Fermi pocket. Without spin-orbit coupling (SOC), the *ab initio* calculations suggest CaAl$_2$Si$_2$ as a system hosting a topological nodal-line setting around the Γ point in the Brillouin zone close to the Fermi level. Once including the SOC, the fragile nodal-line will be gapped and a pair of Dirac points emerge along the high symmetric Γ-A direction locating at the Brillouin zone coordinates (0, 0, $k_z^D \approx \pm 0.278 \times \frac{2\pi}{c}$), which is about 1.22 eV below the Fermi level. The SOC can also induce a topological insulator state along the Γ-A direction with a gap of about 3 meV. The results demonstrate CaAl$_2$Si$_2$ as an excellent platform for the study of novel topological physics with multiple topological states.

†The authors contributed equally to this work.

*Corresponding authors:
wzhao@hit.edu.cn,
gangxu@hust.edu.cn,
guoyf@shanghaitech.edu.cn.

## I. INTRODUCTION

Topological semimetals (TSMs) have been subjected to immense interest because their nontrivial topological band structure could serve as a unique venue for discovering exotic physical properties both in bulk and surface states [1-4], represented by the Dirac fermions [2, 5-8], Weyl fermions [1, 9-16], Majorana fermions [17, 18], and other exotic new fermions beyond Dirac and Weyl fermions [19-23]. The emergence of different topological fermions generally requires protection from specific symmetries and topology of the electronic band structure. For Dirac semimetals (DSMs) with both time-reversal (TR) and space-inversion (SI) symmetries, the fourfold degenerate Dirac point (DP) formed by two degenerate nodes with opposite chirality also requires the protection of additional symmetry from being annihilated or separated, such as a certain crystalline symmetry [2, 6-8]. Once the TR or/and SI is broken, the DP is split into a pair of doubly degenerate Weyl points (WPs), of which the chirality is also symmetry protected [1, 10, 14, 15]. The realization of Dirac and Weyl fermions in solids has bridged with those predicated in high-energy physics. However, the Lorentz invariance is strictly required in high-energy physics whereas is not necessary in solids. When the low-energy excitation breaks the Lorentz invariance, the Dirac/Weyl cones are tilted strongly along a certain momentum direction and the DPs and WPs appear on the boundaries between the hole and electron pockets, thus forming the type-II DSM and WSM [24-27]. The peculiar band topology in the type-II family can produce many exotic phenomena, such as Klein tunneling in momentum space [28], anisotropic electric transport [29], angle-dependent chiral anomaly [30], etc.

Regarding the various nontrivial topological states, it is naturally a question that whether one can combine different topological states into a single material, so that to produce multiple functionalities from the well separated nontrivial topological states, or even the interplay among them. However, due to the different symmetry requirements, the simultaneous realization of different topological states in a single material is very difficult and such materials are in fact very rare. Despite of these difficulties, several materials were theoretically predicated to host multiple topological states and some of them were even experimentally verified. In the noncentrosymmetric cubic B20-type (space group: $P2_13$) CoSi, the chiral crystal symmetry plays an essential role in protecting the unconventional multifold chiral fermions, i.e. the spin-1 chiral fermion carrying a Chern number C = ±2 with threefold band crossing and charge-2 DPs with fourfold degeneracy [19, 20], proved by the very recent angle-resolve photoemission spectroscopy (ARPES) measurements [22, 23]. In the layered transition metal telluride TaIrTe$_4$, in addition to the well separated WPs that were already predicated in the $k$ space above the Fermi level $E_F$, a pair of nodal lines protected by mirror symmetry was also detected by the ARPES measurements [31]. These exciting experimental successes have highly advanced our knowledge on the band topology theory. Theoretically proposed candidates other than CoSi and TaIrTe$_4$ include C$_4$Li [32], the kagome compound Mg$_3$Bi$_2$ [33] and some polar hexagonal *ABC* crystals such as SrHgSn and CaHgSn [34], etc. The nonsymmorphic symmetries in C$_4$Li protected nodal-line could coexist with the type-II DPs, while Mg$_3$Bi$_2$ hosts a type-II nodal-line with the protection of both TR and SI symmetries, and the spin-orbit coupling (SOC) could induce a pair of 3D DPs which are independent of the nodal-line. In SrHgSn and CaHgSn, the crystal point symmetry, especially the sixfold rotation symmetry, protects a pair of band inversion generated DPs setting along the polar rotation axis. Simultaneously, six pairs of WPs originated from inversion symmetry breaking caused by the HgPb layer bulking are in the plane perpendicular to the polar axis and are protected by the combined twofold rotation and TR symmetries. These predictions, however, are still waiting for experimental realization.

In this paper we present studies by means of magnetotransport measurements and *ab initio* calculations on the layered $CaAl_2Si_2$ semimetal crystallized into the structure depicted in Figs. 1(a)-(b) viewed along different orientations. The magnetotransport measurements reveal quantum oscillations of magnetoresistance (MR) and the crucial role of SOC in giving rise to different topological states. The quantum oscillations of MR are tightly related to the electron Fermi pocket of the Fermi surface (FS) located at the M point of the Brillouin zone (BZ). Without the SOC, the hole FS meets a nodal-line near the $E_F$. When the SOC is considered, the nodal-line is gapped and a pair of DPs protected by the $C_3$ symmetry appears along the Γ-A direction together with the SOC induced topological insulator (TI) state.

## II. EXPERIMENTAL

The $CaAl_2Si_2$ crystals were grown by using the self-flux method. Starting materials of Ca (99.95%, aladdin), Al (99.999%, aladdin) and Si (99.9999%, aladdin) blocks were mixed in a molar ratio of 1: 20: 2 and placed into an alumina crucible which was then sealed into a quartz tube in vacuum. The assembly was heated in a furnace up to 1100 °C within 10 hrs, kept at the temperature for 20 hrs, and then slowly cooled down to 750 °C at a temperature decreasing rate of 0.5 °C/hr. The excess Al was removed at this temperature by quickly placing the assembly into a high-speed centrifuge and black crystals with shining surface in a typical dimension of 2.6×2×1.5 mm$^3$, shown by the picture as an inset of Fig. 1(c), were finally left.

The phase and quality examinations of $CaAl_2Si_2$ were performed on a Bruker D8 single crystal X-ray diffractometer (SXRD) with Mo $K_{α1}$ (λ = 0.71073Å) at 298 K. The diffraction pattern could be satisfyingly indexed on the basis of a trigonal structure with the lattice parameters $a = b = 4.137$ Å, $c = 7.131$ Å, $α = β = 90°$ and $γ = 120°$ in the space group *P*-3*m*1 (No. 164), consistent with those reported earlier [35].

The prefect reciprocal space lattice without any other miscellaneous points, seen in Figs. 1(d)-1(f), indicates pure phase and high quality of the crystal used in this study. Paying close attention again to the schematic crystal structure in Fig. 1(a) which is drawn based on the refinement results from the SXRD data, we can find that the Al and Si atoms are arranged in double-corrugated hexagonal layers which are intercalated with Ca in such a way that a layered pile of $-Al_2Si_2-Ca-Al_2Si_2-$ is formed along the $c$-axis. The magnetotransport measurements, including the resistivity and Hall effect measurements, were carried out using a standard Hall bar geometry in commercial DynaCool physical properties measurement system (PPMS) from Quantum Design. We used three crystals from the same batch for each of the resistivity, Hall effect, and magneto-transport measurements. The obtained data of the three crystals closely agree with each other, guaranteeing the reliability of the results.

The first-principles calculations were carried out within the framework of the projector augmented wave (PAW) method [36, 37] and employed the generalized gradient approximation (GGA) [38] with Perdew-Burke-Ernzerhof (PBE) formula [39], as implemented in the Vienna *ab initio* Simulation Package (VASP) [40-42]. A kinetic energy cutoff of 500 eV and a Γ-centered $k$ mesh of 10×10×6 were utilized in all calculations. The energy and force difference criterion were defined as $10^{-6}$ eV and 0.01 eV/Å for self-consistent convergence and structural relaxation. The WANNIER90 package [43-45] was adopted to construct Wannier functions from the first-principles results without an iterative maximal-localization procedure. The WANNIERTOOLS [46] code was used to investigate the topological features of surface state spectra.

## III. RESULTS AND DISCUSSION

The temperature ($T$) dependence of longitudinal resistivity $\rho_{xx}$ measured with $B$ perpendicular to the (001) plane and the electrical current $I$ along the $b$-axis at a magnetic field $B = 0$ T is presented in Fig. 1(c), which displays typical semi-metallic conduction with a residual resistance ratio (*RRR*) $\rho_{xx}(300\text{ K})/\rho_{xx}(2\text{ K})$ of about 9. The application of $B$ = 9 T significantly enhances $\rho_{xx}$ to be somewhat insulating with a plateau behavior at low temperature, which is commonly observed in many topological semimetals [47-50]. Thermal evolution of the $B$ dependent $\rho_{xx}$ measured in the temperature range of 2 − 100 K is summarized in Fig. 2(a) by the main panel. The MR, defined as MR = [$\rho(B) − \rho(0)$]/$\rho(0)$ × 100% in which $\rho(B)$ and $\rho(0)$ represent the resistivity with and without $B$, respectively, shows a crossover from a quadratic-like evolution to linear change, reaching ~1000% at 9 T and 2 K without showing any sign of saturation. Moreover, quantum oscillations in the MR can be observed at $B$ > 6 T below 8 K, which could be seen much clearer by the enlarged view as an inset in Fig. 2(a). This behavior had not been reported in earlier references [51]. The temperature dependence of Hall resistivity $\rho_{xy}$ presented in Fig. 2(b) exhibits clear deviation from the linear change at temperatures below 100 K, indicating that both electron and hole carriers participate the transport. To quantitatively estimate the density and mobility of the carriers in CaAl$_2$Si$_2$, the Hall conductivity, i.e. $\sigma_{xy}$ = - $\rho_{xy}$ / ($\rho_{xy}^2$ + $\rho_{xx}^2$) was fitted by employing the semiclassical two-band model [52, 53]:

$$\sigma_{xy} = \left[ \frac{n_h \mu_h^2}{1 + (\mu_h B)^2} - \frac{n_e \mu_e^2}{1 + (\mu_e B)^2} \right] eB$$

where $n_e$ ($n_h$) denotes the carrier density for the electron (hole), and $\mu_e$ ($\mu_h$) is the mobility of electron (hole), respectively. The fitting results at $T$ = 2 K are presented by the inset of Fig. 2(b), unveiling that the two-band model could nicely describe the Hall conductivity. The fitting yielded temperature dependences of carriers density and mobility for electron and hole carriers are shown in Figs. 2(c) and 2(d), respectively. At 2 K, the carriers density reaches $n_h$ = 7.32 × 10$^{18}$ cm$^{-3}$ and $n_e$ = 4.6 × 10$^{18}$ cm$^{-3}$, revealing that the hole and electron carriers are in fact uncompensated, which could reasonably interpret the smaller MR of CaAl$_2$Si$_2$ as compared with the extremely large MR in other compensated semimetals [54-56]. Though the densities for hole and electron carriers are at the same order of magnitude, the mobility of electron carries is about one order of magnitude larger than that of the hole carriers. Both $n_h$ and $\mu_e$ decrease monotonically with decreasing temperature, whereas $n_e$ shows nearly temperature independent below the temperature of 15 K and $\mu_h$ also reaches a maximal value around the same temperature.

To achieve a further understanding about the electronic band structure and map out the FS, the quantum oscillations of MR were analyzed in more details. After carefully subtracted the smooth background, striking Shubnikov-de Hass (SdH) oscillations in the MR are visible. The SdH oscillations at different temperatures from 2 to 8 K against the reciprocal magnetic field $1/B$ are plotted in Fig. 3(a), which could be well described by the Lifshitz-Kosevich (L-K) formula [57]:

$$\Delta \rho_{xx} \propto R_T R_D \cos\left[2\pi\left(\frac{F}{B} + \varphi\right)\right],$$

where $R_T = 2\pi^2 k_B T/\hbar\omega_c / \sinh(2\pi^2 k_B T/\hbar\omega_c)$, $R_D = \exp(-2\pi^2 k_B T_D/\hbar\omega_c)$, $k_B$ is the Boltzmann constant, $\hbar$ is the Planck's constant, $F$ is the oscillation's frequency, $\varphi$ is the phase shift, $\omega_c = eB/m^*$ is the cyclotron frequency with $m^*$ denoting the effective cyclotron mass, $T_D$ is the Dingle temperature defined by $T_D = \hbar/2\pi k_B \tau_Q$ with $\tau_Q$ being the quantum scattering lifetime. The fast Fourier transform (*FFT*) spectra of the SdH oscillations, depicted in Fig. 3(b), disclose a single fundamental frequency at $F$ = 90 T. The corresponding external cross-sectional area of the FS is $A$ = 0.857 nm$^{-2}$, calculated by using the Onsager relation $F = (\hbar/2\pi e)A$. The effective cyclotron mass $m^*$ at $E_F$ can be estimated from the L-K fitting to the $R_T$, as is shown in Fig. 3(c), giving $m^*$ = 0.417 $m_e$ where $m_e$ denotes the free electron mass. The Fermi wave vector is estimated to be 0.522 nm$^{-1}$ from $k_F = \sqrt{2eF/\hbar}$ and the very large Fermi velocity $v_F$ = 1.45×10$^5$ m s$^{-1}$ is calculated from $v_F = \hbar k_F/m^*$. Fitting to the field dependent amplitudes of the SdH oscillations at 2 K by using the L-K formula, shown in Fig. 3(d), gives the Dingle temperature $T_D$ = 4.89 K and the corresponding quantum scattering lifetime $\tau_Q$ = 2.49 ×10$^{-13}$ s. Furthermore, the quantum mobility $\mu_Q$ = 1048.36 cm$^2$ V$^{-1}$s$^{-1}$ could also be obtained from $\mu_Q = e\tau_Q/m^*$.

The angle-dependent MR is presented in Fig. 4(a) with the measurement geometry shown as the inset. The clear change in angle dependent magnitude of MR exposes the anisotropic nature of the band structure. Fig. 4(b) presents the angle dependent SdH oscillations with a constant offset after subtracting the smooth background, which exhibits clear shift of the peaks with the increase of $\theta$, i.e. the angle between $B$ and the $c$-axis. The $\theta$ dependence of frequency $F$ derived from the SdH oscillations is shown in Fig. 4(c). The frequency changes from 90 T at $\theta = 0°$ (in plane) to 110 T at $\theta = 90°$ (out of plane), clearly unveiling the anisotropy of the Fermi pocket associated with the SdH oscillations. To achieve in-depth insights to the SdH oscillations, the calculated FSs are shown in Fig. 4(d). There are four main sheets of FS, consisting of three hole pockets centered at the $\Gamma$ point and one electron pocket centered at the M point of the BZ. According to the calculation, the hole 2 and 3 pockets can be excluded as the sources producing the SdH oscillations because they are obviously isotropic. The hole pocket 1 has a maximal area in the $ab$ plane, which is not consistent with our experiment data. The electron pocket shows clear anisotropy and the out of plane area is larger than the in-plane one, which is nicely consistent with our analysis on angle-dependent measurements. The SdH oscillations originated from the electron pocket rather than the hole pockets are due to the fact that the mobility of the hole pockets is actually rather small compared with that of the electron pocket, as discussed in the above paragraph. It is known that the quantum lifetime $\tau_Q$ of the electrons is sensitive to all angle scattering processes that could broaden the Landau levels, while the transport lifetime $\tau_T$ is only susceptible to the large angle scattering process. In principle, the $\tau_T/\tau_Q$ ratio is a measure of the relative importance of small angle scattering. According to the Hall effect measurement, the transport lifetime $\tau_T = m^*\mu_e/e = 1.9 \times 10^{-12}$ s, thus yielding the $\tau_T/\tau_Q$ ratio of 8, indicating that the small angle scattering plays a dominated role in the transport. However, this ratio is significantly smaller than that of $Cd_3As_2$[61], consistent with the fact that the carrier mobility is lower.

The calculated band structure in the absence of SOC is shown in Fig. 5(a), indicating that $CaAl_2Si_2$ is a semimetal with an overlap between the valence bands and conduction band. These results are in gratifying agreement with previous electronic structure calculations and electrical resistivity measurements [35, 51, 59]. Note that the hole 1 FS in the shape of a ring torus is quite rare and likely to enclose a nodal line [60]. Along the Γ-K direction, there is a linear band crossing near the $E_F$, seen in Fig. 5(b), whereas a tiny band gap along the Γ-M direction. The band crossing structure is confirmed by additional hybrid functional calculations in the Heyd-Scuseria-Ernzerhof (HSE06) scheme. Fat band analysis shows that there is a band inversion between Al-*s* and Si-*pxy* orbitals at the Γ point (not shown), suggesting nontrivial band topology of CaAl$_2$Si$_2$. Owing to the presence of both TR and SI symmetries, the spinless Hamiltonian is completely real-valued [33, 61]. The band crossing as such does not singly appear but actually persists along a closed path around the Γ point, as schematically plotted in Fig. 5(c). Because of the absence of protection from mirror reflection, glide plane, or screw axes symmetries, the nodal-line in CaAl$_2$Si$_2$ does not exactly lie on the $k_z = 0$ plane and is fragile to the SOC [62]. Fig. 5(d) shows the surface spectrum for the semi-infinite (001) surface, in which the nontrivial drumhead-like surface states, indicated by white arrows in the inset of Fig. 5(d), dispersing outwards from the nodal line are clearly observed. When SOC is considered, the nodal-line in CaAl$_2$Si$_2$ is gapped, as shown in Fig. 5(e).

Another effect of SOC is that it produces a symmetry protected linear band crossing between the bands with $\Delta_4$ and $\Delta_{5+6}$ irreducible representations (IRs) along the Γ-A direction, as shown in Figs. 6(a) and 6(b). The band crossing is fourfold degenerate and protected by $C_3$ rotation symmetry, forming a pair of Dirac points located at BZ coordinates (0, 0, $k_z^D \approx 0.278 \times \frac{2\pi}{c}$) and energy $E - E_F = -1.22$ eV. Note that the band crossing is tilted along Γ-A, while it is straight perpendicular to Γ-K, seen in Fig. 6(c). In addition to topological DSM states, SOC induces hybridization between two $\Delta_4$ bands and opens a gap of about 3 meV, leading to TI states, which resembles that of Fe(Te, Se) [63] and LiFeAs [64]. For both DSM states and TI states, nontrivial topological surface states should exist. The calculated surface states on the (001) surface with SOC is illustrated in Fig. 6(d) to further confirm the topological nature of $CaAl_2Si_2$. For the TI states, the surface shows protected nontrivial Dirac cone type topological surface states inside the SOC gap at the Γ point. For the DSM states, the surface exhibits a gapless linear-dispersion bulk state and a linear Dirac cone surface state.

The nontrivial topological states in $CaAl_2Si_2$ naturally recall into our mind the isostructral topological semimetals $EuCd_2As_2$ and $EuCd_2Sb_2$ [65-68], which belong to the type IV magnetic space group. They were initially proposed to host DSM states, whereas the latter experiments revealed that the peculiar A-type antiferromagnetic structure could break the $C_3$ symmetry along the *c*-axis, and the DP near $E_F$ along the Γ-Z path of BZ is no longer protected by the combined inversion and non-symmorphic time-reversal symmetries, but rather opens tiny band gaps, making the two compounds topological insulators rather than DSMs [69]. In $CaAl_2Si_2$, the absence of magnetism could well reserve the $C_3$ symmetry and hence the DSM state, thus highlighting again the necessary role of certain symmetries in protection the nontrivial topological states.

## IV. SUMMARY

In conclusion, the peculiar electronic band structure of the layered $CaAl_2Si_2$ plays a decisive role in dominating the magnetotransport properties. We have unveiled that $CaAl_2Si_2$ is a topological semimetal hosting intriguing multiple nontrivial topological states that could be tuned by SOC. In the absence of SOC, it holds topological nodal-line structure close to the $E_F$. Once SOC is taken into account, the topological nodal-line will be gapped and a pair of Dirac points, together with the SOC induced topological insulator state, will emerge along the highly symmetric Γ-A direction. These intriguing multiple nontrivial topological states make $CaAl_2Si_2$ an excellent platform for the study of novel topological physics and exploring multiple functionalities.

ACKNOWLEDEMENTS

The authors acknowledge the support by the Natural Science Foundation of Shanghai (Grant No. 17ZR1443300), the National Natural Science Foundation of China (Grant No. 11874264) and the strategic Priority Research Program of Chinese Academy of Sciences (Grant No. XDA18000000). Y.F.G. acknowledges the starting grant of ShanghaiTech University and the Program for Professor of Special Appointment (Shanghai Eastern Scholar). W. Z. is supported by the Shenzhen Peacock Team Plan (Grant No. KQTD20170809110344233), and Bureau of Industry and Information Technology of Shenzhen through the Graphene Manufacturing Innovation Center (Grant No. 201901161514).

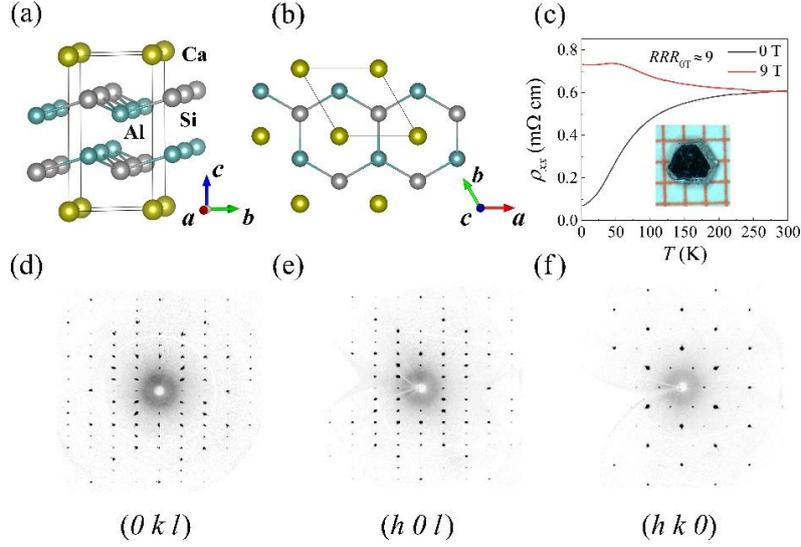

FIG. 1. (a)-(b) Schematic crystal structure of $CaAl_2Si_2$ viewed along different orientations. (c) Temperature dependence of the longitudinal resistivity $\rho_{xx}$. Inset shows an image of a typical single crystal. (d)-(f) Diffraction patterns in the reciprocal space along ($0kl$), ($h0l$), and ($hk0$) directions.

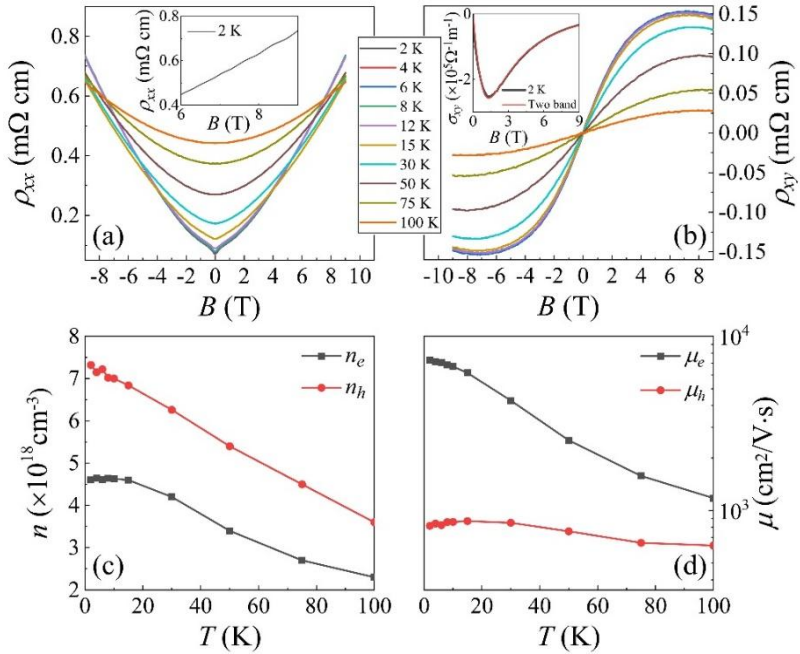

FIG. 2. (a) Longitudinal MR $\rho_{xx}$ versus magnetic field $B$ between $T = 2 - 100$ K. Inset enlarges the oscillation at 2 K. (b) Hall resistivity versus magnetic field $B$ at different temperatures. Inset shows hall conductivity at 2 K and the red line denotes the fitting by the two-band model. (c)-(d) show the temperature dependence of carrier densities and mobilities, respectively.

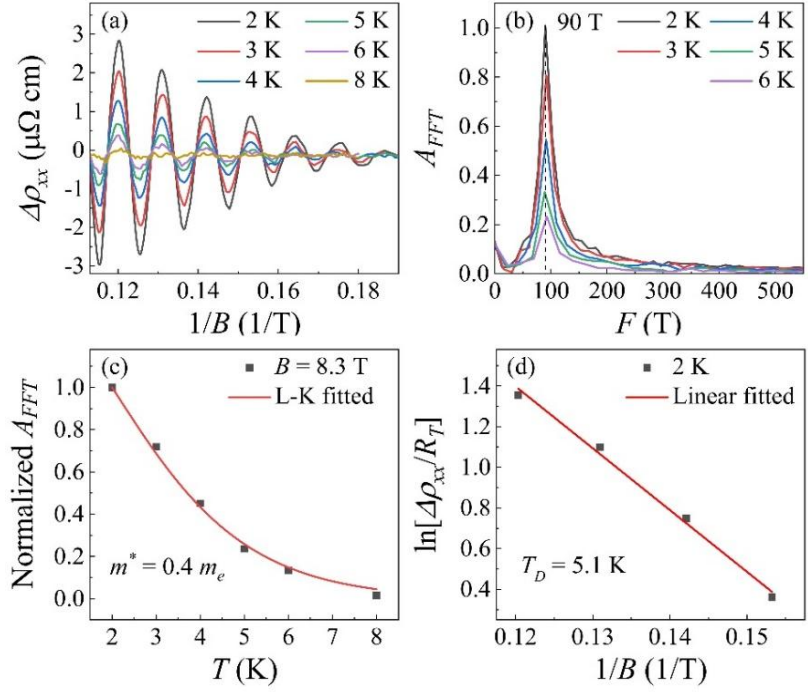

FIG. 3. (a) SdH oscillatory component as a function of $1/B$ after subtracting the smooth MR background. (b) FFT spectra of $\Delta\rho_{xx}$. (c) Temperature dependence of relative amplitudes of the SdH oscillations at $B = 8.3$ T. The solid lines denote the fitting by using the L-K formula. (d) Dingle plot of the SdH oscillations at $T = 2$ K.

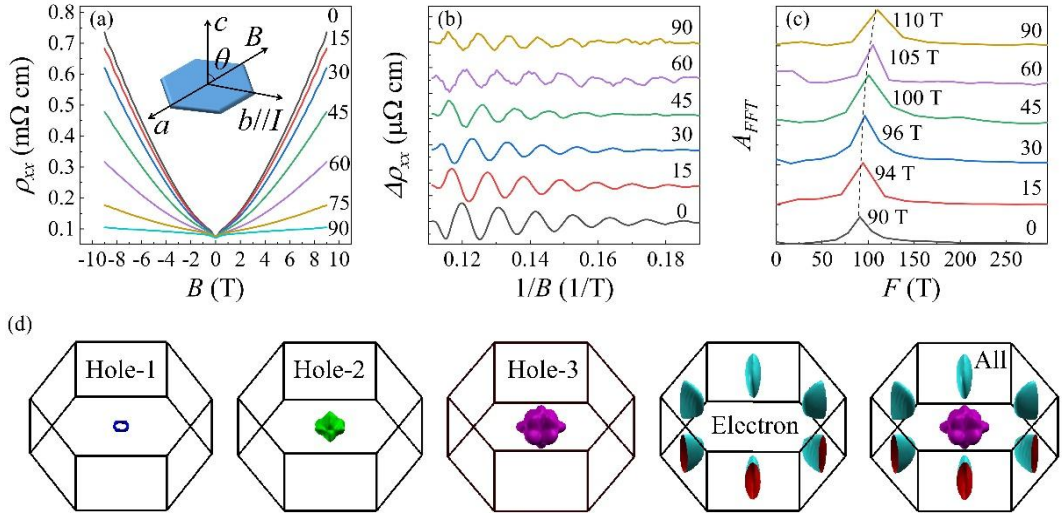

FIG. 4. (a) Longitudinal $\rho_{xx}$ versus magnetic field $B$ at different angles when $T = 2$ K. Inset shows the schematic measurement configuration. (b) SdH oscillatory component as a function of $1/B$ at different angles. (c) FFT spectra of $\Delta\rho_{xx}$ at different angles. (d) Calculated Fermi surfaces of $CaAl_2Si_2$.

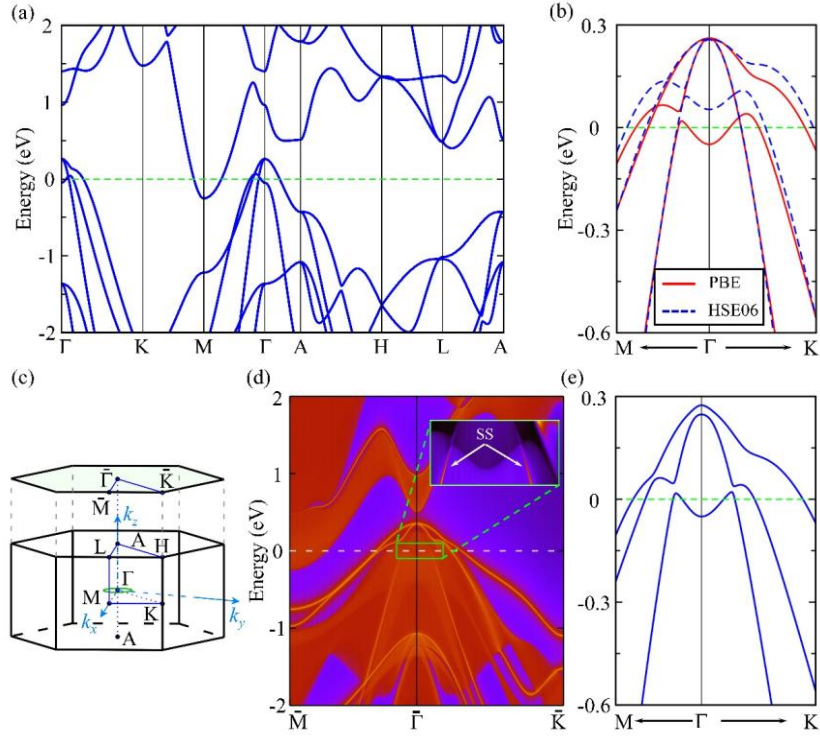

FIG. 5. (a) Electronic band structure of $CaAl_2Si_2$ without SOC. (b) Enlarged band structure along Γ-M and Γ-K paths calculated at PBE and HSE06 levels of theory. (c) Bulk BZ, (001) projected surface BZ and high symmetry points. The nodal-line inside the BZ is illustrated in green. (d) Surface band structure of $CaAl_2Si_2$ on the (001) projected surface. The inset shows the zoom-in view of the solid green box area, where the nontrivial topological surface states situate outside the projected nodal line are pointed out by white arrows and clearly visible. (e) Electronic band structure $CaAl_2Si_2$ with including SOC.

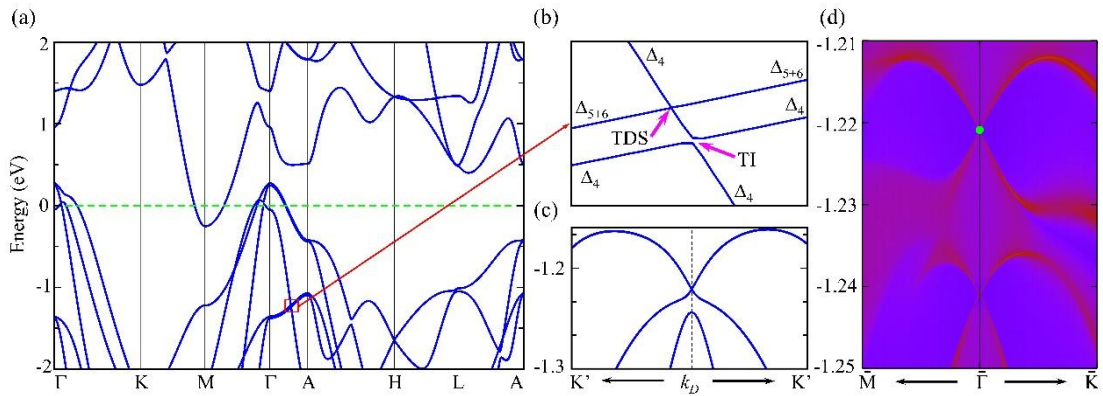

FIG. 6. (a) Electronic band structure of $CaAl_2Si_2$ with considering SOC. (b) Enlarged view of the solid red area in (a). The irreducible representations of selected bands along the high symmetric $k$ are indicated. (c) Electronic band structure in the $k_x$ - $k_y$ plane surrounding the Dirac point. (d) The (001) surface band structure of $CaAl_2Si_2$ in the presence of SOC.